# Title: Constraints from globular cluster pulsars on the magnetic field in the Galactic halo

**Authors:** Federico Abbate[1,2]*, Andrea Possenti[2,3], Caterina Tiburzi[4,5], Ewan Barr[4], Willem van Straten[6], Alessandro Ridolfi[2,4], Paulo Freire[4]

**Affiliations:**

[1]Dipartimento di Fisica `G. Occhialini', Università degli Studi Milano - Bicocca, Piazza della Scienza 3, I-20126 Milano, Italy.

[2]INAF - Osservatorio Astronomico di Cagliari, Via della Scienza 5, I-09047 Selargius (CA), Italy.

[3]Università di Cagliari, Dip. di Fisica, S.P. Monserrato-Sestu Km 0,700 - I-09042 Monserrato (CA), Italy.

[4]Max-Planck-Institut für Radioastronomie, Auf dem Hügel 69, D-53121 Bonn, Germany.

[5]Fakultät für Physik, Universität Bielefeld, Postfach 100131, 33501 Bielefeld, Germany.

[6]Institute for Radio Astronomy & Space Research, Auckland University of Technology, Private Bag 92006, Auckland 1142, New Zealand.

*Correspondence to: f.abbate@campus.unimib.it

**The Galactic magnetic field plays an important role in the evolution of the Galaxy, but its small-scale behaviour is still poorly known. It is also unknown whether it permeates the halo of the Galaxy or not. By using observations of pulsars in the halo globular cluster 47 Tucanae, we probed the Galactic magnetic field at arcsecond scales for the first time and discovered an unexpected large gradient in the component of the magnetic field parallel to the line of sight. This gradient is aligned with a direction perpendicular to the Galactic disk and could be explained by magnetic fields amplified to some 60 μG within the globular cluster. This scenario supports the existence of a magnetized outflow that extends from the Galactic disk to the halo and interacts with the studied globular cluster.**

The Galactic magnetic field has important effects on the evolution of the Galaxy, by affecting star formation, propagation of cosmic rays and by regulating Galactic winds. However, its origin and precise structure is poorly known (*1,2*). The Galactic magnetic field in the disk is thought to be made of an ordered component at large scales and a random component at small scales with a turbulent nature (*3*). The magnetic field in the halo that surrounds the Galaxy is also poorly known (*4*); it is usually thought to be predominantly azimuthal (*5*) but a component perpendicular to the Galactic disk has been suggested (*6,7*) and found some support (*8-10*). A possible origin of this component is a magnetized outflow from the disk (*11*) whose existence is supported by radio observations (*12*) and by diffuse Galactic X-ray observations (*13*).

Pulsar observations have led to important results on the determination of the large-scale structure of the Galactic magnetic field (*14-17*). The radiation of pulsars is typically highly polarized and, as it travels through an ionized and magnetized medium, the polarization angle rotates by an amount proportional to the column density of the ionized gas, the parallel component of the magnetic field and the square of the wavelength of the radiation (*18*). This effect is called Faraday rotation and is quantified by the rotation measure (RM). Thanks to the periodic nature of the pulsar radiation and its broadband emission, it is also possible to directly measure dispersion in the ionized interstellar medium, which is also proportional to the square of the wavelength and the column density of the ionized gas, via a quantity known as dispersion measure (DM). After estimating both dispersion and rotation measures, it is possible to isolate the effects of the magnetic field along the line of sight and study it directly. Such studies are typically limited by the large angular separation of pulsars, usually more than a degree (*16*), and uncertainty in pulsar distances. In this regard, globular clusters (GCs) can help us since they typically host large numbers of pulsars with small angular separation, from arcseconds to arcminutes, and their distances are known with great accuracy. In comparison, the smallest angular scales of the magnetic field analysed in studies of extragalactic RMs go down to only ~10 arcminutes (*19,20*). Furthermore, a large fraction of GCs is located in the Galactic halo. Thus, observations of GC pulsars can substantially contribute to the study of the intensity and geometry of magnetic fields in the halo.

An excellent GC for this kind of analysis is 47 Tucanae (also known as NGC 104, hereafter 47 Tuc). This GC is at a distance of 4.5 kpc and located in the Galactic halo close to the South Celestial Pole, in a region where few magnetization studies have been conducted. It has a Galactic latitude of −45 deg, a height from the disc of 3.2 kpc and the distance from the Galactic centre projected on the disc is 6.8 kpc. It contains 25 known pulsars (*21-25*) that are, with one exception, located within 1 arcminute from the centre (*26,27*). The pulsars from this cluster have been studied extensively to find the first evidence of ionized gas inside a GC (*28*), test different models of gas distribution and estimate the three-dimensional positions of the pulsars (*29*). The ionized gas is found to have a uniform distribution with a density of $0.23 \pm 0.05$ cm$^{-3}$ near the centre of the cluster, where the pulsars are located. This gas is thought to originate from the winds of evolved stars and is constantly ejected from the cluster (*30*). We report the results of polarimetric observations of these pulsars.

**Observations and Results**

The observations were made with the Parkes radio telescope in Australia between April 2014 and March 2015 using the central beam of the Parkes Multi-Beam Receiver (*31*) with a central frequency of 1382 MHz and a bandwidth of 400 MHz. The CASPSR backend was used. The data were coherently dedispersed, acquired in full Stokes mode with a sampling time of 32 μs and successively calibrated in flux and in polarization (see Methods section). Of the 25 pulsars, only 18 were detected and for only 13 of them it was possible to measure the RM. The polarization profiles of the detected pulsars are shown in Supplementary Figures 1-3.

The measured RMs, polarization percentages, and the values of DM are reported in Table 1. The average value of RM is ~ 13 rad m$^{-2}$ and the standard deviation is ~ 13 rad m$^{-2}$. Initial results indicated a correlation between the RMs and the spatial distribution of the pulsars. To investigate this, we performed a fit using a Bayesian maximum likelihood algorithm (see Methods section) and found that the dataset is well described by a linear gradient (Fig. 1 and 2). The parameters of the fit and the best-fitting values are the magnitude of the gradient, $m = -0.77 \pm 0.06$ rad m$^{-2}$ arcsec$^{-1}$, the inclination angle of the gradient measured from Celestial North to East, $\theta = 30 \pm 2$ deg, and the value of RM at the centre of the cluster, $RM_0 = 20 \pm 1$ rad m$^{-2}$. The reduced chi-square of the fit is 0.7 with 10 degrees of freedom. We checked if a random distribution of RMs could reproduce the results by applying a Bayesian model selection algorithm (see Methods section); we found that there is only a 0.04 per cent probability of a random distribution to obtain a better fit to the data. This means that the observed gradient is not due to random fluctuations at a 3.5-sigma level.

**Table 1.** Measured properties of the pulsars detected in the observations. We report the percentage of linear polarization, L, and value of RM. For reference we also report the DM of the pulsars (*27*). For some pulsars the RM was not measurable either because of low signal-to-noise ratio or low polarization fraction. In these cases we write a dash in the corresponding cell. The errors represent the 68% confidence interval.

| Name | L (per cent) | RM (rad m$^{-2}$) | DM (pc cm$^{-3}$) |
|---|---|---|---|
| C | 15.4 ± 0.4 | 33.4 ± 2.4 | 24.600 ± 0.001 |
| D | 13.3 ± 1.8 | 11.8 ± 11.8 | 24.732 ± 0.001 |
| E | 27.4 ± 1.1 | 27.2 ± 2.2 | 24.236 ± 0.001 |
| F | 18.7 ± 1.8 | 17.4 ± 8.1 | 24.382 ± 0.001 |
| G | 44.1 ± 2.4 | 12.2 ± 6.8 | 24.436 ± 0.003 |
| H | 13.1 ± 4.7 | - | 24.369 ± 0.003 |
| I | 46.0 ± 1.7 | 4.9 ± 5.8 | 24.429 ± 0.004 |
| J | 5.9 ± 0.2 | -9.1 ± 3.1 | 24.588 ± 0.003 |
| L | 55.9 ± 10.2 | 18.7 ± 11.0 | 24.400 ±0.004 |
| M | 12.3 ± 3.7 | - | 24.433 ±0.006 |
| N | 62.7 ± 5.4 | -0.4 ±5.6 | 24.573 ± 0.005 |
| O | 9.8 ± 1.6 | 23.8 ± 17.0 | 24.356 ± 0.002 |
| Q | 23.5 ± 3.4 | -9.1 ± 9.9 | 24.266 ± 0.003 |
| R | 14.4 ± 4.9 | - | 24.361 ± 0.003 |
| S | 24.3 ± 3.7 | - | 24.376 ± 0.002 |
| T | 15.7 ± 1.9 | 12.2 ± 12.7 | 24.393± 0.016 |
| U | 0.0 ± 5.1 | - | 24.337 ± 0.002 |
| Y | 45.6 ± 2.7 | 24.5 ± 3.5 | 24.468 ± 0.002 |

**Analysis and Discussion**

If we assume that the RM distribution is caused by differences in the electron density along the line of sight, the latter differences would be reflected in the measured DMs. Assuming that the magnetic field is uncorrelated with the electron density, the relation between magnetic field, RM and DM is given by (*18*):

$$\langle B_\parallel \rangle = 1.23\ \mu G \left(\frac{RM}{\text{rad m}^{-2}}\right) \left(\frac{DM}{\text{pc cm}^{-3}}\right)^{-1}$$

Where $B_\parallel$ is the component of the magnetic field parallel to the line of sight. Assuming a constant magnetic field of ~ 5 µG (a typical value for the Galactic magnetic field in the disk (*2*)), an RM spread of ~ 40 rad m$^{-2}$ would require a DM range of order ~ 10 pc cm$^{-3}$, much larger than the observed maximum of 0.6 pc cm$^{-3}$. Therefore, the RM spread cannot be ascribed to fluctuations in the electron density.

Alternatively, the observed broad range of values of RMs could arise from differences in the parallel component of the magnetic field along different lines of sight over the angular scales probed by the pulsars. The magnetic field responsible for that could in turn be located either in the Galactic disk along the line of sight, in the GC itself or in the Galactic halo. In the following we explore these possibilities separately.

At sub-arcminute scales, the small-scale magnetic field in the Galactic disk is thought to follow the electron density distribution described consistently by the Kolmogorov theory of turbulence (*32,3*). We tested whether the observed RMs could be described by a turbulent field through the study of the RM structure function, but it was inconclusive. The description and results of this analysis are reported in the Methods section and plotted in Supplementary Figure 6. While the standard deviation of RM is comparable with what has been found in previous studies at larger angular separations at similar Galactic latitudes (*19,20*), turbulent fluctuations are not expected to show strong correlations with a specific spatial direction. Furthermore, the fluctuations of DM due to the interstellar medium in the Galactic disk, after removing the contribution from the GC gas, have a standard deviation of ~ 0.1 pc cm$^{-3}$ (*29*).

If the observed RM fluctuations (which have a standard deviation of ~ 13 rad m$^{-2}$) were due to only the fluctuations of dispersion in the Galactic disk, then the parallel component of the magnetic field would be ~ 150 µG, which is almost two orders of magnitude greater than the expected value.

We now explore the second hypothesis that the RM variations on the small scales covered by 47 Tuc pulsars are due to an ordered magnetic field located inside the GC. To obtain an estimate of the required strength of this magnetic field we start from the definition of the RM (*33*):

$$\mathrm{RM} = 0.81 \int n_e(l)\, B_\parallel(l)\, dl \ \mathrm{rad\ m^{-2}}$$

Where $n_e \sim 0.23$ cm$^{-3}$ (*29*) is the number density of free electrons, considered constant over the central parts of the cluster, $B_\parallel$ is the component of the magnetic field parallel to the line of sight and the integral is extended over the central region of the cluster where the pulsars and the gas are located, about 2 pc (*29*). Assuming a constant strength of the magnetic field, the parallel component will have different values depending on where the field lines are pointing. Using the above equation, we find that to explain the observed RM difference of ~ 40 rad m$^{-2}$ we need a difference in the parallel magnetic field component of ~ 100 µG. This value must be compared with the equipartition value of the magnetic field measured using previously determined parameters of the cluster (*29*) which is ~ 4 µG, making this picture very unlikely.

A third option invokes the combined effects of the interaction between the wind released by the Galactic disk and the movement of the GC. In the following we show that in this case a shock front arises providing the needed amplification of the magnetic field transported by the wind. We note that the inclination angle of the gradient, $\theta$, is compatible at $2\sigma$ with the direction perpendicular to the Galactic disk, which measured with our conventions would be ~ 26 deg. The geometry of the field lines (as resulting from our discussion below) is shown in Fig. 3. In this model, the field lines are perpendicular to the Galactic disk and reach the cluster forming an angle $\theta_H$ measured from Celestial North to East in the plane of the sky and an angle $\varphi$ with respect to the line of sight. The angle $\varphi$ corresponds to the absolute value of the Galactic colatitude of the cluster. In our case it is expected to be 45 deg. The Galactic wind is ejected from the Galactic disk with a velocity of ~ 200 km s$^{-1}$ (*13*) while the GC is moving towards the Galactic disk with a velocity of ~ 45 km s$^{-1}$ (*34*). The total velocity of 47 Tuc with respect to the wind in the direction orthogonal to the Galactic disk is ~ 245 km s$^{-1}$. This velocity is both superalfvenic ($v_a$ ~ 175 km s$^{-1}$) and supersonic ($c_s$ ~ 170 km s$^{-1}$) with respect to the wind; therefore, a shock forms in front of the cluster. Crossing the shock, the magnetic field lines are bent and acquire a component perpendicular to the motion of the cluster (see discussion in the Methods section). The geometry of the shock interface in presented in Extended Data 1. This perpendicular component is compressed together with the gas and amplified owing to magnetic flux conservation.

In order to estimate the factor of amplification, we first estimate the properties of the gas after entering the shock interface by applying ideal magnetohydrodynamics, modified for a collisionless astrophysical plasma (see Methods section). In order to achieve equilibrium with the gas present in the cluster, the wind is compressed further. This compression of the gas strongly amplifies the perpendicular component of the magnetic field. Simple considerations (see

Methods section) show that, in this situation, the magnetic field is capable of reaching strengths of up to ∼ 80 μG.

However, at first approximation, we can schematically model the magnetic field with a semi-circular geometry and note that the direction in which the field lines rotate changes with respect to the position where they cross the globular cluster. Field lines are circular in the half of the cluster facing the Galactic wind and straight in the other half. Field lines above the axis H shown in Fig. 3 rotate clockwise while field lines under it rotate counter-clockwise.

We perform a fit of this model leaving as free parameters the strength of the magnetic field, $B$, the direction of the Galactic wind in the plane of the sky, $\theta_H$, the Galactic foreground contribution to RM, $RM_{0,H}$ and the inclination angle with respect to the line of sight, $\varphi$. The best fit to the data is shown in Fig. 4 where the RM predicted by the model is plotted versus the measured RM (see Methods section for details on the fit). The reduced chi-square is 1.0 with 9 degrees of freedom while the best-fitting value of the inclination angle $\theta_H$ is $36 \pm 4$ degrees which is compatible at 2.5σ with the direction from the GC to the Galactic disk. The best-fitting value of the foreground contribution to RM is $23 \pm 2$ rad m$^{-2}$ which is compatible with the values of $30 \pm 8$ rad m$^{-2}$ (*35*) and of $16 \pm 10$ rad m$^{-2}$ (*36*) estimated for this region in extragalactic RM studies. The best-fitting value of the angle φ is $63 \pm 10$ deg, compatible at 2σ with the expected value of 45 deg. In different models (*8*), the Galactic outflow is not perfectly perpendicular to the Galactic disk so even a marginal compatibility in the direction is acceptable. The intensity of the magnetic field required to explain the observed gradient in RM is $66 \pm 11$ $\mu$G again compatible with the estimated ∼ 80 μG.

Thus, we conclude that the source of the gradient in RM is likely a magnetic field located inside 47 Tuc and amplified by the interaction with a Galactic wind. The required strength means that the magnetic field plays an important role in the internal dynamics of the cluster and must therefore be studied in more details with magnetohydrodynamical simulations. The model can be better tested using observations made with the MeerKAT radio telescope (*37*) in South Africa, which will be able to determine the RMs of a greater number of pulsars with higher precision. If a Galactic wind permeating the halo is responsible for the observed magnetic field, then similar effects should also be visible in other GCs populating the Galactic halo. Observing these GCs will be helpful in identifying the best model for the halo magnetic field (*6*).

**Corresponding Author:** All correspondence and request for materials should be addressed to F. A.

**Acknowledgments:** We acknowledge the help of Andrew Jameson in performing the observations. We would like to thank Dr. Kuo Liu, Dr. Sui Ann Mao and Prof. Michael Kramer for useful comments. The Parkes radio telescope is part of the Australia Telescope, which is funded by the Commonwealth of Australia for operation as a National Facility managed by the Commonwealth Scientific and Industrial Research Organisation. Authors are indebted to the communities behind the multiple open-source software packages on which this work depended. This research made use of Astropy, a community-developed core Python package for Astronomy. **Funding:** F.A., A.P., A.R. acknowledge the support from the Ministero degli Affari Esteri della Cooperazione Internazionale - Direzione Generale per la Promozione del Sistema Paese - Progetto di Grande Rilevanza ZA18GR02. Part of this work has also been funded using resources from the research grant "*iPeska*" (P.I. A. Possenti) funded under the INAF national call Prin-SKA/CTA approved with the Presidential Decree 70/2016.

**Author contributions:** F.A. calibrated the data, estimated the RM, created the magnetic field models, performed the statistical analysis and compiled the manuscript. A.P. conceived and supervised the project and revised the manuscript. C.T. helped in the calibration and RM estimation process and revised the manuscript. E.B. provided access to the data, pre-analysed the observations and revised the manuscript. W.v.S. provided crucial help in the polarization calibration process and revised the manuscript. A.R. and P.F. shared the latest timing results and revised the manuscript. A.R. also helped in the production of the polarization profiles shown in Supplementary Figures 1-3.

**Competing interests:** Authors declare no competing interests.

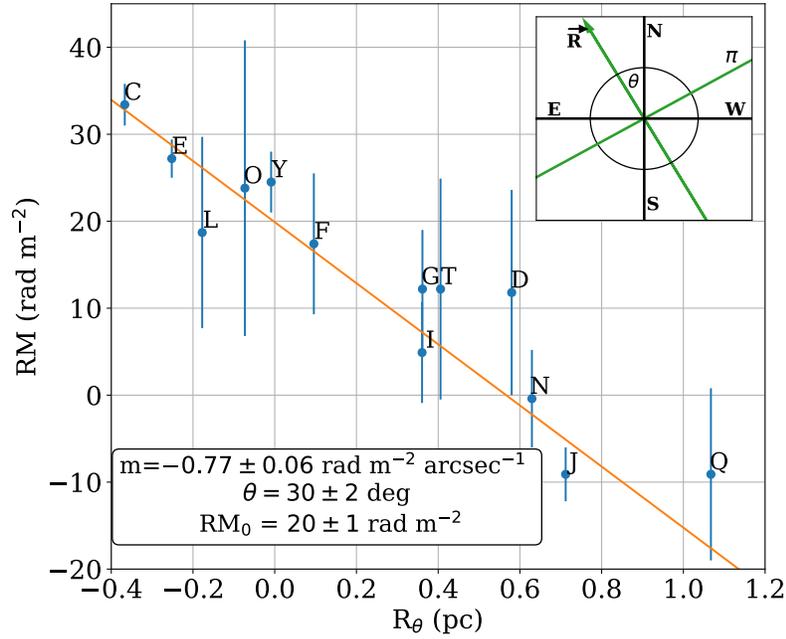

**Fig. 1.** Detected gradient of RM as a function off the pulsar positions. RM values as a function of the projected distance of each pulsar from the plane π perpendicular to the axis $\vec{R}$ and intersecting $\vec{R}$ at the centre of the cluster (*38*). $\vec{R}$ is oriented with an angle of 30 degrees measured from the North direction to East. The orange line is the best line fit through the data. The panel in the top right shows the direction of the gradient. The box in the bottom left contains the best-fitting parameters. The error bars are at 1σ.

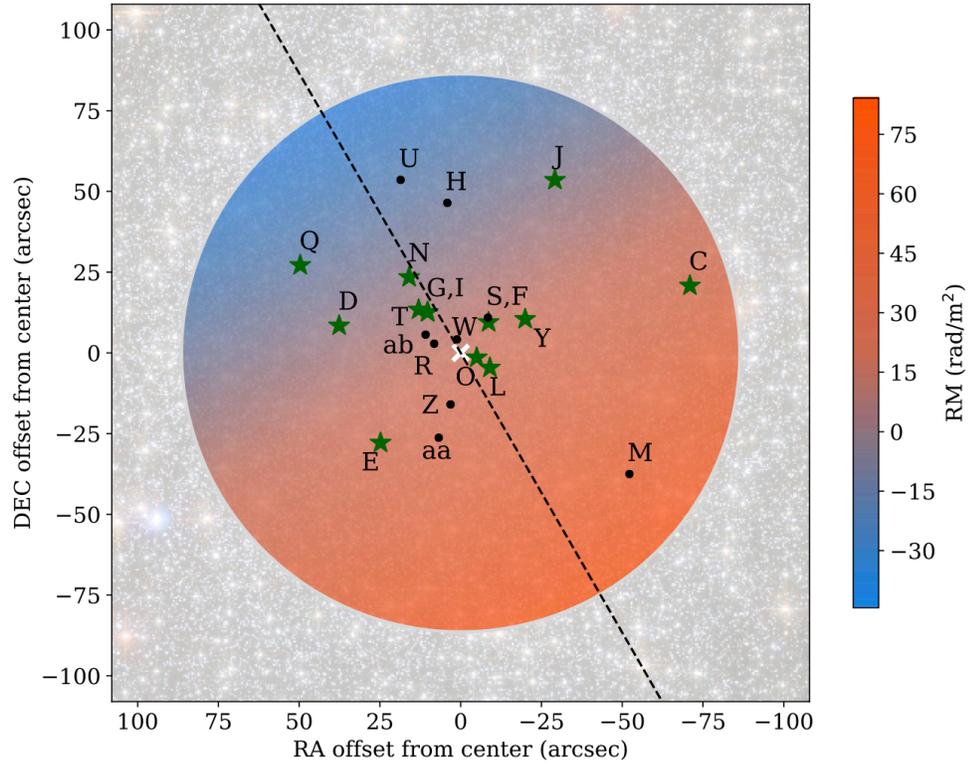

**Fig. 2.** Map showing the gradient in RM and the distribution of pulsars in 47 Tuc. The colour scale shows the RM values predicted by the best-fitting linear gradient model presented in Fig. 1. The pulsars for which a measurement of RM was possible are represented by green stars while the pulsars with no measured RM are represented by black dots. The dashed line shows the direction of the $\vec{R}$ axis in Fig. 1. A white cross marks the optical centre of the cluster (*38*). Credits of background image: NASA, ESA, and the Hubble Heritage (STScI/AURA)-ESA/Hubble Collaboration. Acknowledgment: J. Mack (STScI) and G. Piotto (University of Padova, Italy).

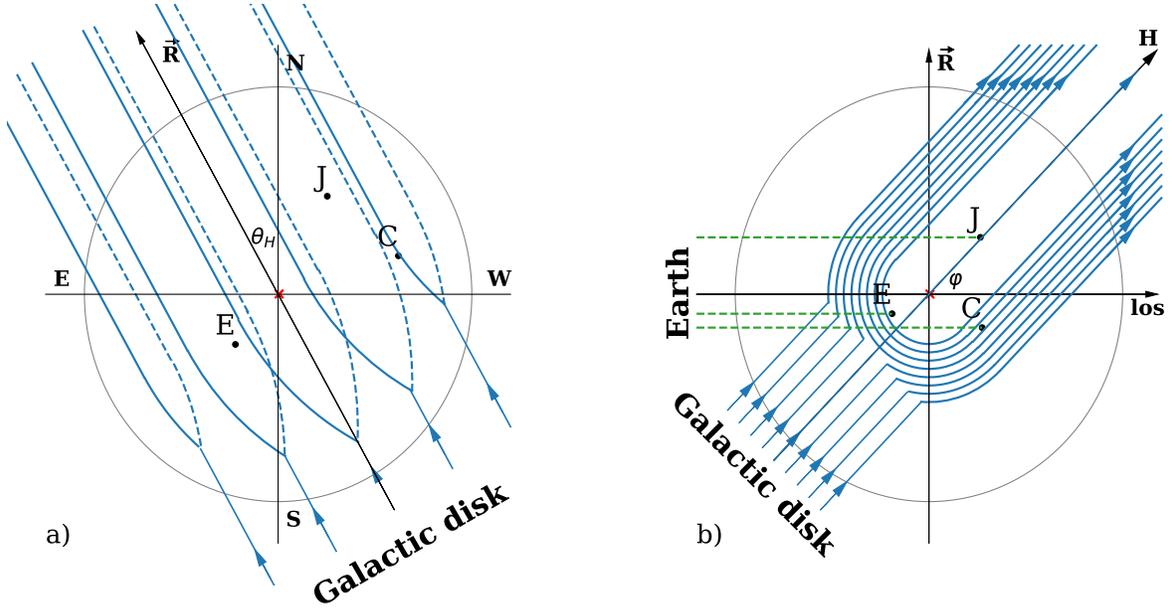

**Fig. 3.** Magnetic field lines in the interaction between a Galactic wind and 47 Tuc. The magnetic field lines, thought to reach 47 Tuc along a direction perpendicular to the Galactic disk, are bent by the presence and the motion of the gas in the cluster. The centre of the cluster is marked with a red cross. Panel a) shows the projection on the plane of the sky in celestial coordinates. The angle $\theta_H$ is compatible with the angle $\theta$ defined in the linear model in Fig. 1. Panel b) shows the same model in the plane defined by $\vec{R}$ and the line of sight. Three pulsars with precise RM values and known line of sight position (*29*), 47 Tuc C, E, and J, are also shown. The dashed green lines in panel b) are the lines of sight to the pulsars. The value of RM is influenced only by the component of the magnetic field along these lines of sight. This model of the magnetic field structure inside 47 Tuc produces a RM contribution that is negative for pulsar 47 Tuc J and positive for pulsars 47 Tuc C and 47 Tuc E.

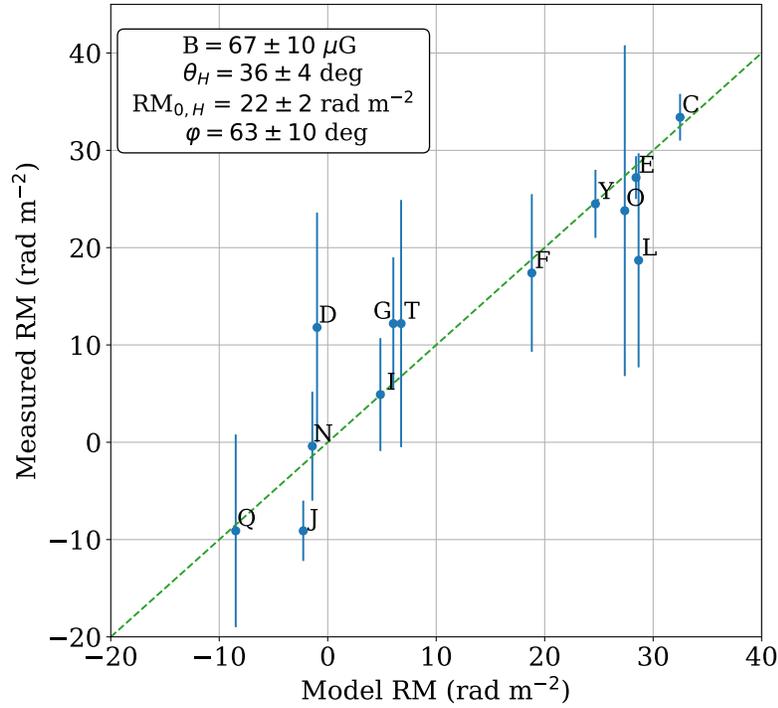

**Fig. 4.** Best fit of the RM in the Galactic wind model. The Galactic wind is assumed to follow the geometry described in the text and in Fig. 3. The plot shows the measured RM versus the RM predicted by the best fit. The best-fitting values for the parameters are shown in the box. The error bars are at 1σ. For a perfect fit, all points should fall on the dashed green line, which represents the identity function.

**Materials and Methods**

Calibration procedure of pulsar observations and RM estimation

The observations were recorded in baseband mode and processed offline separately for each pulsar. The data were reduced and calibrated with the PSRCHIVE software package (*39,40*). Each observation was first cleaned from Radio Frequency Interference (RFI) both in frequency and time. To remove strong RFI at the edges of the observing band, the total bandwidth was reduced from 400 MHz to 312 MHz. The data were calibrated in flux using averaged observations of Hydra A and calibrated in polarization using the Measurement Equation Template Matching (*41*) technique. This method incorporates measurements of a noise diode and observations of pulsar J0437-4715 to derive corrected Stokes parameters. The pulsars were coherently dedispersed to entirely remove the effects of pulse dispersion caused by the ionised gas along the line of sight and were folded according to the best available timing solutions (*26,27*). Different observations were summed together to obtain integrated profiles with high signal-to-noise ratio (SNR). The pulsars in 47 Tuc are heavily affected by scintillation so only the best observations for each pulsar were used to produce the integrated profiles. The number of profile bins for each pulsar was reduced in order to smooth the profile.

Two methods were used to measure the RM. The first method is implemented in the RMFIT routine of PSRCHIVE. the polarization position angle (PA) is the same across the whole observing band and that frequency-dependent PA rotation along the line of sight depolarizes the profile. The routine applies different RM values, calculates the linearly polarized component of the total signal and returns the RM which leads to the highest level of linear polarization. This method works better in the presence of high SNR and high polarization. For most of the pulsar data available for this work the routine was not capable of calculating a precise value of RM. The second method used (*16,42*) is more accurate and consists in measuring the PA from the Stokes parameters at different wavelengths and searching for the RM that best fits the defining equation:

$$\Delta\Psi_{PA} = RM\, c^2 \frac{1}{f^2}$$

where $\Delta\Psi_{PA}$ is the rotation of the PA, $c$ is the speed of light and $f$ is the observing frequency.

We proceeded by selecting the on-pulse, linearly polarized region, and an off-pulse region dominated by noise. The number of frequency channels was reduced to four to increase the SNR in each of them. Each channel has a width of 78 MHz with centre frequencies: 1264, 1342, 1421 and 1499 MHz. For each channel we measured the average PA along the pulse with the formula:

$$PA = \frac{1}{2}\tan^{-1}\left(\frac{\sum_{i=n_{start}}^{n_{end}} U_i}{\sum_{i=n_{start}}^{n_{end}} Q_i}\right)$$

where $U_i$ and $Q_i$ are the Stokes parameters $U$ and $Q$ for the $i^{th}$ profile bin and $n_{start}$ and $n_{end}$ are the start and the end of the on-pulse region.

To calculate the error on the PAs, we first measured the total linear polarization as:

$$L_{meas} = \sqrt{\left(\sum_{i=n_{start}}^{n_{end}} U_i\right)^2 + \left(\sum_{i=n_{start}}^{n_{end}} Q_i\right)^2}$$

Since $U$ and $Q$ are affected by noise and $L_{meas}$ is a positive definite quantity, it is positively biased. To remove the bias we calculated the signal-to-noise ratio of the measured linear polarization $p_0 = L_{meas}/(rms(I)\sqrt{n_{pulse}})$, where $rms(I)$ is the off-pulse root mean square of the total intensity profile and $n_{pulse}$ is the number of bins in the pulse region (*43*). The true value of *L* is:

$$L_{\text{true}} = \begin{cases} 0.0 & \text{if } p_0 < 2.0 \\ \sqrt{L_{\text{meas}}^2 - \left(rms(I)\sqrt{n_{pulse}}\right)^2} & \text{otherwise} \end{cases}$$

This is the best correction when $p_0 > 0.7$ (*44*). The error of the PAs can be determined by measuring the signal-to-noise ratio of the unbiased linear polarization $P_0 = L_{\text{true}}/(rms(I)\sqrt{n_{pulse}})$; if $P_0 > 10$, the underlying distribution can be described by a Gaussian and the error is: $\sigma_{PA} = \frac{1}{2P_0}$ (*45*). If instead $P_0$ is lower, the assumption of a normal distribution is not valid. In this case we need to use the following normalized distribution (*46*):

$$G(\text{PA - PA}_{\text{true}};\ P_0) = \frac{1}{\sqrt{\pi}} \left( \frac{1}{\sqrt{\pi}} + \eta_0 e^{\eta_0^2}[1 + \text{erf}(\eta_0)] \right) e^{-\frac{P_0^2}{2}}$$

where $\text{PA}_{\text{true}}$ is the measured value of PA, $\eta_0 = (P_0/\sqrt{2})\cos 2(\text{PA - PA}_{\text{true}})$ and erf is the Gaussian error function. This distribution is not analytically integrable and to calculate the error we need to numerically integrate this distribution between $\pm\sigma_{PA}$ in order to obtain 0.683. As we are only looking at the pulsed emission, we know that all of the polarized emission is coming from the pulsar. Therefore, in the pulsar data we only expect to find a single Faraday component. We performed a least-square fit to find the RM according to the formula:

$$\text{PA}(f) = \text{RM}\, c^2 \frac{1}{f^2} + \text{PA}_\infty$$

where $\text{PA}_\infty$ is the value of PA at very high frequency. Since the number of frequency channels has been reduced to 4 it is not possible to test possible variations from the formula reported. Some pulsars are known to present variations of RM as a function of rotational phase (*47,48*). To face this problem, where there is enough SNR, we divide the pulse into regions and perform a simultaneous fit with a single value of RM. The value of $PA_\infty$ can be different for every region and does not influence the value of RM resulting from the fit, so we leave it as a free parameter. This way we recover an average value which is close to the real RM. Two examples of these fits are reported in Supplementary Figure 4.

This approach is also useful when dealing with pulsars with a multi-peaked profile and for pulsars that show large variations of PA across the pulse phase. In pulsars with multi-peaked profile summing all of the polarization information from the pulse together will add a considerable amount of noise which will increase the uncertainty of the measure. In pulsars that show large variations of PA summing together signals over a large rotational phase will lead to a depolarisation of the signal and larger uncertainties. In both cases, this approach leads to a more precise estimate of RM.

Since the error on the PAs is not always Gaussian, the least-square fitting algorithm does not always return the correct uncertainty on RM. To measure it in a more statistically sound way we perform a Monte Carlo simulation. In this simulation we created synthetic profiles of the real pulsars for the $U$ and $Q$ polarization in four frequency bands with the same polarization and added a random Gaussian noise with $\text{rms}(I)$ as standard deviation. We calculated the value of RM with the procedure described above for 1000 times and showed that the results followed a Gaussian distribution. We took the standard deviation of the Gaussian of the simulated results as representative of the error on the RM.

This technique of estimating RM can be affected by the n-π ambiguity. Measuring an angle using the arctangent function always returns a value between -π/2 and π/2. If the RM value of the pulsar is such that the PA rotation in the observed frequency band is higher than π/2, the measured values of PA will be wrong by a factor π and the fit will not return the correct value of RM. To solve this problem, we measured the differences between PAs of neighbouring frequency channels and if these differences are higher than $\pi/2$ we add or subtract $\pi$ in order to correct the PA. This method has problems if the RM is so high that between two adjacent frequency bins the rotation induced is higher than $\pi/2$. This happens if RM is higher than 247 rad m$^{-2}$ or lower than -247 rad m$^{-2}$. The n-$\pi$ ambiguity does not affect the routine RMFIT so, for the pulsars for which RMFIT was successful, we checked if the results were compatible. For the pulsars for which RMFIT was not successful, we assumed the true value of RM to be as close as possible to the other pulsars.

If both methods were successful, the results were combined with a weighted average to obtain the reported value. If only one method returned an estimate of RM, that value is taken. Another method that can be used to measure RM is RM synthesis (*49,50*). This method was not used in the present analysis because it does not work as well as the line fitting method in the case of low-SNR sources (see Figure 9 in *49*).

The profiles of the pulsars corrected for the measured RM and summed at all frequencies are shown in Fig. S1-S3. In these figures we also plot the PAs as a function of pulse phase only for the phase bins in which the linear polarization has been detected over $3\sigma$. The plotted values of PAs are corrected for the effect of the RM by only considering the value $PA_\infty$ estimated in the equation above. These values are not frequency-dependent. For the pulsars with no measured RM, no correction has been applied.

A source of error for the RM that was neglected is the ionospheric contribution. This is typically between 0.5 and 3 rad m$^{-2}$ and shows strong diurnal variations (*51,52*). Since our measured errors are usually larger, applying this correction would not change the results significantly.

The linear polarization percentages are measured with:

$$L \text{ (per cent)} = \frac{1}{n_{\text{pulse}}} \sum_{i=n_{\text{start}}}^{n_{\text{end}}} L_{\text{true},i} \times \frac{100}{S_0}$$

where $S_0 = \frac{1}{n_{\text{pulse}}} \sum_{i=n_{\text{start}}}^{n_{\text{end}}} I_i$ is the flux density, Ii is the total intensity of the bin I, and $L_{\text{true},i}$ is the values of linear polarization of the bin $i$.

Bayesian analysis

The statistical analysis used throughout the paper is based on a Bayesian algorithm that uses flat prior on all parameters and, assuming there are N instances of the quantity being fitted for, called X, maximizes a Gaussian log-likelihood of the form:

$$L \propto \sum_{i}^{N} -\frac{1}{2\sigma_{X,i}} \left(X_{\text{meas},i} - X_{\text{model},i}\right)^2$$

where $X_{\text{meas},i}$ is the measured value of X on the $i^{\text{th}}$ instance, $X_{\text{model},i}$ is the prediction of the model and $\sigma_{X,i}$ is the standard deviation of the measure. The maximization is performed using the EMCEE Python package (version 3.0.2) (*53*), which implements a Markov Chain Monte Carlo algorithm and returns the best-fitting parameters for the desired model.

Details of the linear RM fit

In the case of the linear fit shown in Fig. 1, the model used to fit the RM as a function of the position of the pulsars is $\text{RM}_{\text{lin}}(i|m, \theta, \text{RM}_0) = mR_{i,\theta} + \text{RM}_0$, where $R_{i,\theta}$ is the distance of each pulsar relative to the cluster centre and projected along the axis in the plane of the sky with an inclination angle of $\theta$ (measured from North to East) and is defined as $R_{i,\theta} = \text{RA}_i \sin\theta + \text{DEC}_i \cos\theta$, where $\text{RA}_i$ and $\text{DEC}_i$ are the right ascension and declination of each pulsar after subtracting the right ascension and declination of the cluster centre. This code makes use of the Astropy package (version 2.0.9) (*54, 55*) The geometry of this model is shown in Supplementary Figure 5. The free parameters are the magnitude of the gradient $m$, the inclination angle $\theta$, and the value of RM at the cluster centre, $\text{RM}_0$.

We also checked if the linear correlation found could be replicated by a random distribution of RM. To do so we repeated this analysis 10000 times extracting the RMs from a uniform random distribution between -15 and 35 rad m$^{-2}$ making sure the maximum difference between RMs is close to the one observed and with the same uncertainties as the measured values. We calculated the probability of randomly reproducing a fit which is comparable to the measured one by using a Bayesian model selection algorithm based on the Bayes factor. If the ratio of the Bayes factors of the two fits is within 0.01 and 100, then the fits are considered comparable. This happens in 1.5% of the cases, but the Bayes factor of the randomly extracted data exceeds the one measured with the observed data in only 4 cases out of 10,000 trials, Therefore, the quality of the measured fit cannot be exceeded by random data at the 3.5$\sigma$ level.

Details of the RM structure function spectral index estimation

To check if the observed RMs could be described by a turbulent magnetic field, we calculated the RM structure function, defined as:

$$D_{RM}(\delta\theta) = \langle [RM(\theta) - RM(\theta + \delta\theta)]^2 \rangle_\theta$$

The average (indicated by the angular brackets) is computed between all pairs of pulsars with an angular separation of $\delta\theta$ on the sky. We first measured the square of the RM difference for each pair of pulsars and averaged them over 7 equally spaced bins each containing around 10 pairs. The spectral index is estimated by performing a straight-line fit through the data in logarithmic units using the algorithm described above. The model used was: $\log D_{RM}(i|\alpha,k) = \alpha \log l_i + k$, where $i$ is the bin number, $l_i$ is the center position of $i^{th}$ bin, $\alpha$ is the spectral index and $k$ is a normalization. The free parameters of the fit are $\alpha$ and $k$. The measured RM structure function is shown in Supplementary Figure 6. The value of the best-fitting normalization parameter $k$ is $1.3^{+0.7}_{-0.9}$. and the best-fitting spectral index is $\alpha = 0.8^{+0.6}_{-0.4}$ which is lower than the value of 5/3 predicted by the turbulent theory (*32,56*) but still consistent with what has been observed in RM studies at larger angular separations (*19*). This analysis therefore cannot completely rule out a turbulent magnetic field. The first measure of an RM structure function using the pulsars in a globular cluster was performed by Anna Ho (private communications).

Details of the Galactic wind model

The values of RM are linked to physical quantities like the strength of the magnetic field and the gas density through the following equation (*16*):

$$RM = \frac{e^3}{2\pi m_e^2 c^4} \int_0^d n_e(x) \, B_\parallel(x) \, dx$$

Here $e$ is the electron charge, $m_e$ is the electron mass, $c$ is the speed of light, $d$ is the distance travelled by the light along the line of sight, $n_e$ is the electron density and $B_\parallel$ is the component of the magnetic field parallel to the line of sight. Assuming that the magnetic field responsible for the gradient is located in the GC and that the Galactic magnetic field contribution to RM is constant for all pulsars, we only compute the integral within the GC. The electron density inside the cluster is assumed to be constant at a value of $n_e = 0.23$ cm$^{-3}$ (*29*).

The model of the interaction between the Galactic wind and the globular cluster is composed of semi-circular magnetic field lines of constant strength centred around an axis passing through the centre of the cluster. In the half of the cluster that is facing the Galactic wind, the field lines are circular, while in the other half they are straight. The geometry of the field lines is shown in Fig. 3.

In order to derive the analytical expression of the magnetic field we first measure the projection of the position of the pulsars on axis R as was done in the linear model: $R_{i,\theta_H} = \mathrm{RA}_i \sin\theta_H + \mathrm{DEC}_i \cos\theta_H$. In the half of the plane R-los, defined in Fig. 3b, that is facing the Galactic wind, the magnetic field is oriented in a circular direction. If we call B the direction of the magnetic field, then it can be described by:

$$\hat{B} = -\frac{R_{i,\theta_H}}{\sqrt{x_i^2 + R_{i,\theta_H}^2}} \hat{x} + \frac{x_i}{\sqrt{x_i^2 + R_{i,\theta_H}^2}} \hat{R}$$

Where $x_i$ is the distance of each pulsar along the line of sight from a plane parallel to the sky passing through the centre of the cluster and $\hat{B}, \hat{x}$ and $\hat{R}$ are the unit vectors respectively along the magnetic field, line of sight and axis $\vec{R}$. In the computation of RM only the component parallel to the line of sight is relevant so we will only consider the component directed along $\hat{x}$. The direction of the magnetic field changes in the two quarters divided by the axis H in panel b) of Fig. 3. Above this axis the field lines will be rotating clockwise while under it the field lines will be rotating counter-clockwise. In the half of the cluster that is facing away from the wind, the field lines are straight and only the component along the line of sight enters the equation. The transition between circular and linear field lines happens along the line passing through the centre perpendicular to the direction H that can be parametrized as $x_{H,\perp} = -R_{i,\theta_H} \tan \varphi$. Integrating along the line of sight to estimate the contribution to RM, we need to consider this sign change. The change occurs at the position along the line of sight $x_H = \frac{R_{i,\theta_H}}{\tan \varphi}$. The total RM for each pulsar assumes the for

$$\mathrm{RM}_{\mathrm{wind}}(i|B, \theta_H, \mathrm{RM}_{0,H}, \varphi) =$$

$$= \begin{cases} -\frac{e^3}{2\pi m_e^2 c^4} n_e B \int_{x_{max}}^{x_i} \frac{R_{i,\theta_H}}{\sqrt{x^2 + R_{i,\theta_H}^2}} dx + \mathrm{RM}_{0,H} & x_i < x_H \text{ and } x_i < x_{H,\perp} \\ -\frac{e^3}{2\pi m_e^2 c^4} n_e B \left( \int_{x_{max}}^{x_H} \frac{R_{i,\theta_H}}{\sqrt{x^2 + R_{i,\theta_H}^2}} dx - \int_{x_H}^{x_i} \frac{R_{i,\theta_H}}{\sqrt{x^2 + R_{i,\theta_H}^2}} dx \right) + \mathrm{RM}_{0,H} & x_i > x_H \text{ and } x_i < x_{H,\perp} \\ -\frac{e^3}{2\pi m_e^2 c^4} n_e B \left( \int_{x_{max}}^{x_{H,\perp}} \frac{R_{i,\theta_H}}{\sqrt{x^2 + R_{i,\theta_H}^2}} dx + \int_{x_{H,\perp}}^{x_i} \cos \varphi \, dx \right) + \mathrm{RM}_{0,H} & x_i > 0 \text{ and } x_i > x_{H,\perp} \\ -\frac{e^3}{2\pi m_e^2 c^4} n_e B \left( \int_{x_{max}}^{x_H} \frac{R_{i,\theta_H}}{\sqrt{x^2 + R_{i,\theta_H}^2}} dx - \int_{x_H}^{x_{H,\perp}} \frac{R_{i,\theta_H}}{\sqrt{x^2 + R_{i,\theta_H}^2}} dx + \int_{x_{H,\perp}}^{x_i} \cos \varphi \, dx \right) + \mathrm{RM}_{0,H} & x_i < 0 \text{ and } x_i > x_{H,\perp} \end{cases}$$

Here $B$ is the strength of the magnetic field, assumed to be constant and $\mathrm{RM}_{0,H}$ is the Galactic contribution to RM. The value of $x_{max}$ is set to 10 pc, and we verified that the quality of the fit does not depend strongly on this quantity. The minus sign is added because the line of sight component, $x$, is positive when the magnetic field points away from Earth while RM must be negative.

Solving the integrals, we find the equations of the predicted RM as a function of the physical parameters:

$$\text{RM}_{\text{wind}}(i|B, \theta_H, \text{RM}_0, \varphi) =$$

$$= \begin{cases} \frac{e^3}{2\pi m_e^2 c^4} n_e B R_{i,\theta_H} \log\left(\sqrt{x_{max}^2 + R_{i,\theta_H}^2} + x_{max}\right) - \\ \quad - \frac{e^3}{2\pi m_e^2 c^4} n_e B R_{i,\theta_H} \log\left(\sqrt{x_i^2 + R_{i,\theta_H}^2} + x_i\right) + \text{RM}_{0,H}, & x_i < x_H \text{ and } x_i < x_{H,\perp} \\ \\ \frac{e^3}{2\pi m_e^2 c^4} n_e B R_{i,\theta_H} \left[\log\left(\sqrt{x_{max}^2 + R_{i,\theta_H}^2} + x_{max}\right) - 2\log\left(\sqrt{x_H^2 + R_{i,\theta_H}^2} + x_H\right)\right] + \\ \quad + \frac{e^3}{2\pi m_e^2 c^4} n_e B R_{i,\theta_H} \log\left(\sqrt{x_i^2 + R_{i,\theta_H}^2} + x_i\right) + \text{RM}_{0,H}, & x_i > x_H \text{ and } x_i < x_{H,\perp} \\ \\ \frac{e^3}{2\pi m_e^2 c^4} n_e B \left[-\cos\varphi\, (x_i - x_{H,\perp}) + R_{i,\theta_H} \log\left(\sqrt{x_{max}^2 + R_{i,\theta_H}^2} + x_{max}\right)\right] - \\ \quad - \frac{e^3}{2\pi m_e^2 c^4} n_e B R_{i,\theta_H} \log\left(\sqrt{x_{H,\perp}^2 + R_{i,\theta_H}^2} + x_{H,\perp}\right) + \text{RM}_{0,H}, & x_i > 0 \text{ and } x_i > x_{H,\perp} \\ \\ \frac{e^3}{2\pi m_e^2 c^4} n_e B \left[-\cos\varphi\, (x_i - x_{H,\perp}) + R_{i,\theta_H} \log\left(\sqrt{x_{max}^2 + R_{i,\theta_H}^2} + x_{max}\right) - 2R_{i,\theta_H} \log\left(\sqrt{x_H^2 + R_{i,\theta_H}^2} + x_H\right)\right] + \\ \quad + \frac{e^3}{2\pi m_e^2 c^4} n_e B R_{i,\theta_H} \log\left(\sqrt{x_{H,\perp}^2 + R_{i,\theta_H}^2} + x_{H,\perp}\right) + \text{RM}_{0,H}, & x_i < 0 \text{ and } x_i > x_{H,\perp} \end{cases}$$

The free parameters of this model are the strength of the field, $B$, the inclination angle of the axis in the plane of the sky, $\theta_H$, the foreground contribution to RM, $\text{RM}_{0,H}$ and the inclination angle with respect to the line of sight, $\varphi$ that appears in the definition of $x_H$.

Magnetohydrodynamic shock equations

The globular cluster and its internal gas are moving with a speed that is both supersonic and superalfvenic with respect to the Galactic wind, so the development of a shock interface is warranted. When crossing the shock front, the magnetic field in the wind naturally acquires a component perpendicular to the direction of motion which can be compressed and thus amplified. The geometry of the shock is shown in Fig. S7.

To test if this scenario is applicable and could generate a magnetic field comparable with the one observed, we appeal to basic MHD shock equations (*57*). We first move in the frame of reference of the shock so that the upstream and unperturbed material (defined with a subscript 1) is moving towards the shock with a speed of $U_1$, a density of $\rho_1$ and a magnetic field of $B_1$. If the magnetic field is aligned with the direction of $U_1$, the shock can generate a component perpendicular to $B_1$ in the downstream region (defined with a subscript 2) if $U_1 > c_a$, where $c_a$ is the Alfven speed. In our case $U_1 \sim 250$ km s$^{-1}$ while $c_a = \frac{B_1}{\sqrt{4\pi n_1 m_p}} \sim 175$ km s$^{-1}$, where $B_1 \sim 3$ µG, $n_1 \sim 1.4 \times 10^{-3}$ cm$^{-3}$, and $m_p$ is the proton mass, have been derived by scaling the value estimated for the wind (*13*) at the position of the cluster.

As the gas crosses the shock interface it is compressed, its direction changes away from the normal of the shock and any (even very small) perpendicular component of the magnetic field is enhanced according to the following equations (*57*):

$$\frac{n_2}{n_1} = \frac{U_{\|1}}{U_{\|2}} = X_0$$

$$\frac{B_{\perp 2}}{B_{\perp 1}} = \frac{(U_1^2 - c_a^2)X_0}{U_1^2 - X_0 c_a^2}$$

In this equation $U_{\|1}$ and $U_{\|2}$ are the component parallel to the shock normal of the velocities upstream and downstream, $B_{\perp 2}$ and $B_{\perp 1}$ are the perpendicular components to the shock normal in the two regions and $X_0$ is a factor that must be $1 < X_0 \leq \frac{U_1^2}{c_a^2} \sim 2$. The component on the x-axis of the magnetic field is conserved through the shock so the direction of the magnetic field changes according to:

$$\frac{B_{\perp 2}}{B_{\|2}} = \tan\theta_2 = \frac{(U_1^2 - c_a^2) X_0}{U_1^2 - X_0 c_a^2} \tan\theta_1$$

In this equation $\theta_2$ is the angle between $B_2$ and the shock normal and $\theta_1$ is the angle between $B_1$ and the shock normal. Assuming $X_0 \sim 1.5$ (the central value of the possible interval) we find $n_2 \sim 2 \times 10^{-3}$ cm$^{-3}$, $U_{x2} \sim 170$ km s$^{-1}$ and $B_{\perp 2} \sim 3 B_{\perp 1}$. For example, if a magnetic field line enters the shock interface at an angle of 5° it will come out at ~14° with a perpendicular component of ~ 0.8 µG.

The density of the plasma inside the cluster is ~ 0.23 cm$^{-3}$, derived from the electron density (*28*). If the magnetic field is to penetrate in the central regions of the cluster, it must be compressed of a factor of ~ 100. This additional compression further amplifies the perpendicular component of the magnetic field which acquires a circular geometry in the half of the cluster that is facing the shock.

To estimate how much the perpendicular component of the magnetic field can be amplified by this type of compression, take a cube that is compressed only on one side, called l, with the magnetic field aligned along a direction perpendicular to the compression. Mass conservation implies that $nAl$ is constant, where A is the surface area of the face that remains constant. On the other hand, magnetic flux conservation implies that the magnetic flux that crosses the surface of the solid remains constant, in this situation the surface area has a linear dependence with the length of the side that is changing, l, so we have that *Bl* is constant. Therefore, the magnetic field should grow linearly with the density.

Even magnetic field lines that have a perpendicular component of only 0.8 μG (as in the case of a field line that enters the shock with an angle of 5°) are capable of reaching a value of ∼ 80 μG which is even higher than the best-fit magnetic field strength that parameterizes our model of the observed RM gradient. This shows that this mechanism is able to reach the required strengths of magnetic fields even if the Galactic wind is less magnetized than we assumed.

This compression leaves the parallel component of the magnetic field constant and thus bends the field lines further in a semi-circular shape as depicted in Extended Data 1.

Data availability

The data that support the plots within this paper and other findings of this study are available from the corresponding author upon reasonable request.

Code availability

The custom codes used for the analysis and described in the methods section are provided as supplementary information.

**Methods references:**

39. Hotan A. W., van Straten W., Manchester R. N. PSRCHIVE and PSRFITS: An Open Approach to Radio Pulsar Data Storage and Analysis, *Publ. Astron. Soc. Australia*, **21,** 302-309 (2004).

**Extended Data 1.**

Diagram of the shock forming in front of the globular cluster. The shock is cause by the superalfvenic motion of the cluster in the frame of the Galactic wind. The globular cluster (not in scale) is the dashed circle, the thick black line is the shock front and the blue lines are the magnetic field lines. The quantities denoted with the subscript 1 are the velocity, density and magnetic field of the gas in the upstream region, while the quantities denoted with the subscript 2 are the same in the downstream region. The density of the gas in the cluster is denoted by $n_{GC}$.

# Supplementary Information for the paper:
# Constraints from globular cluster pulsars on the magnetic field in the Galactic halo

**Authors:** Federico Abbate[1,2]*, Andrea Possenti[2,3], Caterina Tiburzi[4,5], Ewan Barr[4], Willem van Straten[6], Alessandro Ridolfi[2,4], Paulo Freire[4]

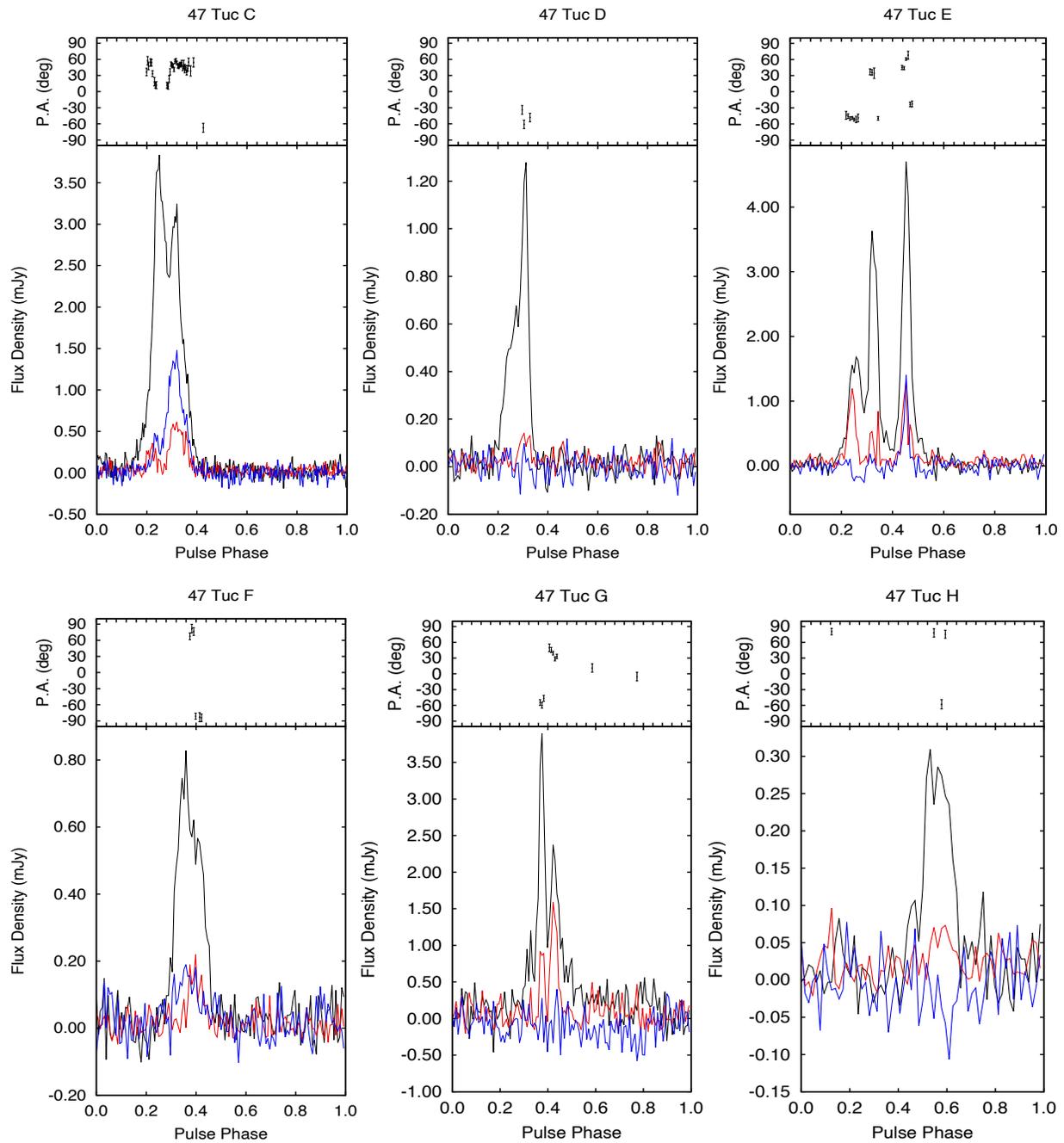

**Supplementary Figure 1.**
Flux calibrated and RM-corrected polarization profiles of the pulsars 47 Tuc C, D, E, F, G, H. All frequencies are summed together after correcting for the interstellar dispersion. The top panel of each plot shows the polarization position angle (PA) variation from the celestial north as a function of pulse phase. The PAs are plotted only if the linear polarization is detected at more than $3\sigma$. The bottom plot shows the flux density of the integrated profile. The black line is the total intensity, the red line is the linear polarization and the blue line is the circular polarization. The centre frequency of the observations is 1382 MHz.

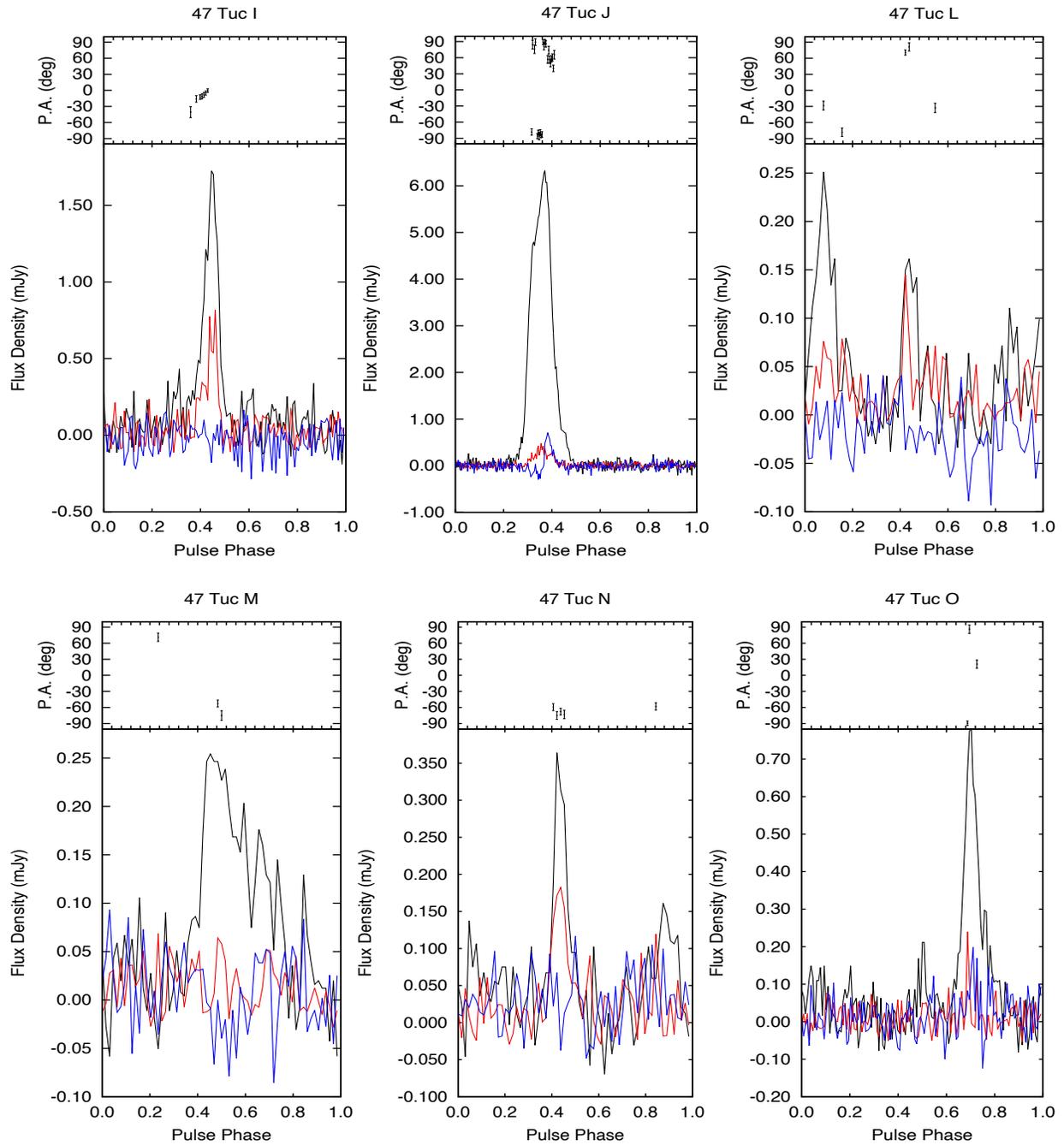

**Supplementary Figure 2.**
Flux calibrated and RM-corrected polarization profiles of the pulsars 47 Tuc I, J, L, M, N and O. All frequencies are summed together after correcting for the interstellar dispersion. The top panel of each plot shows the polarization position angle (PA) variation from the celestial north as a function of pulse phase. The PAs are plotted only if the linear polarization is detected at more than 3σ. The bottom plot shows the flux density of the integrated profile. The black line is the total intensity, the red line is the linear polarization and the blue line is the circular polarization. The centre frequency of the observations is 1382 MHz.

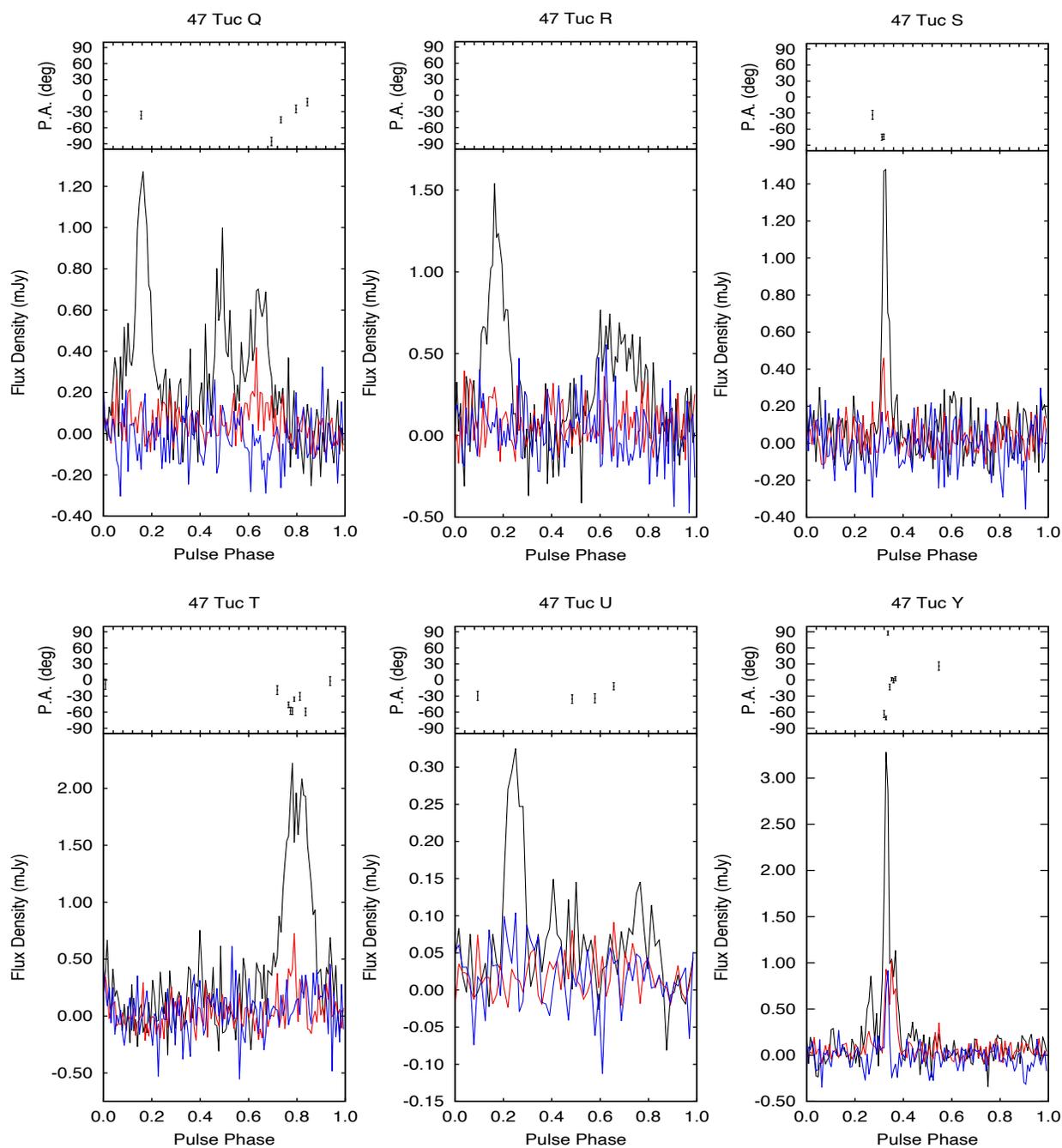

**Supplementary Figure 3.**
Flux calibrated and RM-corrected polarization profiles of the pulsars 47 Tuc Q, R, S, T, U and Y. All frequencies are summed together after correcting for the interstellar dispersion. The top panel of each plot shows the polarization position angle (PA) variation from the celestial north as a function of pulse phase. The PAs are plotted only if the linear polarization is detected at more than 3σ. The bottom plot shows the flux density of the integrated profile. The black line is the total intensity, the red line is the linear polarization and the blue line is the circular polarization. The centre frequency of the observations is 1382 MHz.

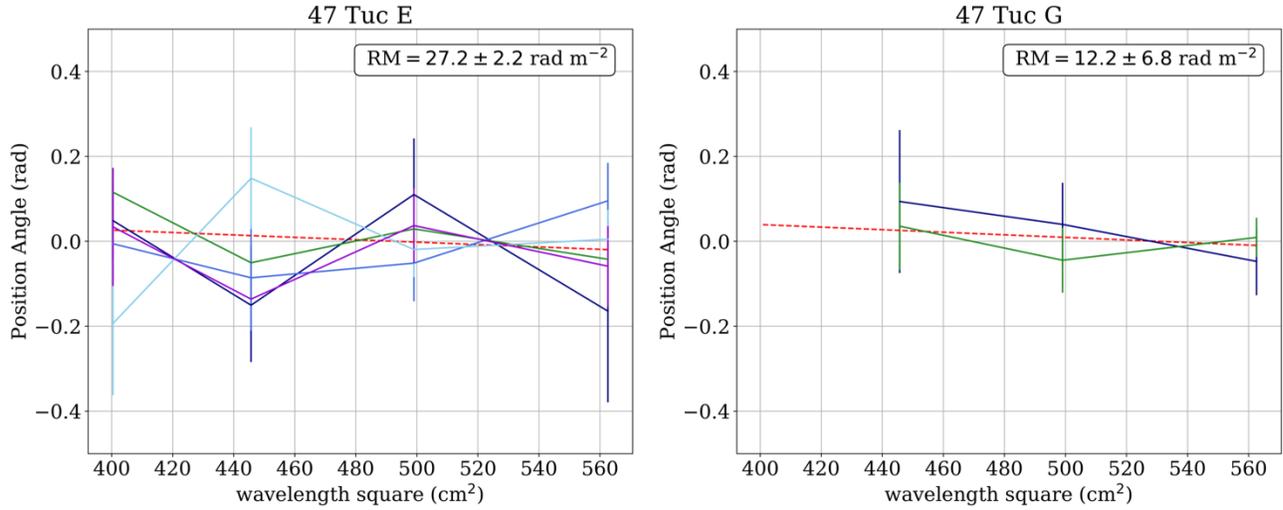

**Supplementary Figure 4.**
Examples of two fits of RM for the pulsars 47 Tuc E and G. Both are examples of simultaneous fits over different regions of pulse longitude. In the case of 47 Tuc E we divided the pulse in 5 regions and performed a fit for a single RM for all regions. To facilitate visual inspection of the fit quality, the mean PA was subtracted from each region. The red dashed line shows the best fit. The data have been previously corrected assuming a RM of 30 rad m$^{-2}$ so a straight line with a slope of zero in the plot corresponds to an RM value of 30 rad m$^{-2}$. For 47 Tuc G we divided the pulse longitude in two regions corresponding to the two peaks of the profile in Fig. S1. The data have been corrected assuming an RM of 15 rad m$^{-2}$. In both regions the PAs corresponding to the first wavelength squared bin have been removed because the linear polarization is too weak ($p_0 < 2.0$).

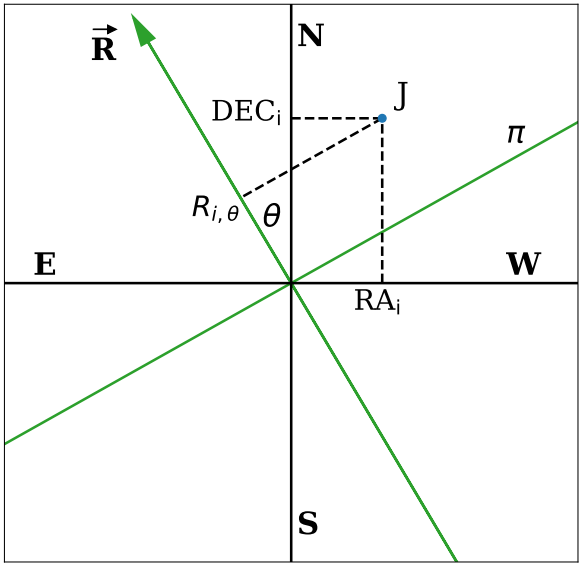

**Supplementary Figure 5.**
Geometry of the linear model as shown in Fig. 1. The pulsar J is shown as an example.

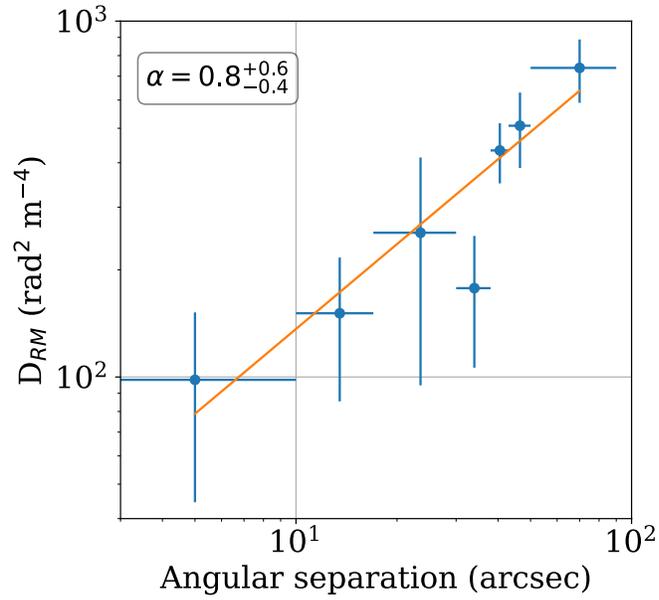

**Supplementary Figure 6.**

Rotation Measure structure function of the pulsars in 47 Tuc. The orange line is the best-fitting power law with an index of $\alpha = 0.8^{+0.6}_{-0.4}$. The reduced chi-square of the fit is 1.2 with 5 degrees of freedom. The error bars on $D_{RM}$ are at $1\sigma$.